\documentclass[prc,onecolumn,notitlepage,floatfix,tightenlines,12pt]{revtex4-1}

\usepackage{graphicx}
\usepackage{color}
\usepackage{transparent}
\usepackage{amsmath,amssymb,bm}
\usepackage{amsfonts}
\usepackage{tensor}
\usepackage{dcolumn}

\usepackage{todonotes}
  \def\nuc#1#2{\relax\ifmmode{}^{#1}{\protect\text{#2}}\else${}^{#1}$#2\fi}
  \def\itnuc#1#2{\setbox\@tempboxa=\hbox{\scriptsize\it #1}
    \def\@tempa{{}^{\box\@tempboxa}\!\protect\text{\it #2}}\relax
    \ifmmode \@tempa \else $\@tempa$\fi}
  \newcommand{\nm}{\ensuremath{N_\mathrm{max}}}
  \newcommand{\ho}{\ensuremath{\hbar \Omega}}
  \newcommand{\lsrg}{\ensuremath{\Lambda_\mathrm{SRG}}}
  
  \newcommand{\nn}{\ensuremath{N\!N}}
  \newcommand{\nnn}{\ensuremath{3N\mathrm{F}}}
  \newcommand{\co}{(Color online)} 

\newcommand{\vertiii}[1]{{\left\vert\kern-0.25ex\left\vert\kern-0.25ex\left\vert #1 
    \right\vert\kern-0.25ex\right\vert\kern-0.25ex\right\vert}}

\newcommand{\braket}[2]{\left \langle #1 \vphantom{#2} \right|
 \left. #2 \vphantom{#1} \right\rangle} 
\newcommand{\rrmatrixel}[3]{\left \langle #1 \vertiii{
 #2 } #3\right \rangle} 

\newcommand{\sexj}[6]
{ \begin{Bmatrix}
  #1 & #2 & #3 \\
  #4 & #5 & #6
\end{Bmatrix}
}

\newcommand{\nioj}[9]{ \begin{Bmatrix}
  #1 & #2 & #3 \\
  #4 & #5 & #6 \\
  #7 & #8 & #9
\end{Bmatrix}}

\begin{document}

\title{Microscopic description of translationally-invariant $\mathrm{core}+N+N$ overlap functions}

\author{D.~S\"a\"af} \affiliation{Department of Fundamental Physics,
  Chalmers University of Technology, SE-412 96 G\"oteborg, Sweden}

\author{C.~Forss\'en} \email{christian.forssen@chalmers.se}
\affiliation{Department of Fundamental Physics,
  Chalmers University of Technology, SE-412 96 G\"oteborg, Sweden}

\begin{abstract} 
  We derive expressions for $\mathrm{core}+N+N$ overlap integrals
  starting from microscopic wave functions obtained in the \emph{ab
    initio} no-core shell model. These overlap integrals correspond to
  three-body channel form factors and can be used to investigate the
  clustering of many-body systems into a core plus two nucleons. We
  consider the case when the composite system and the core are described
  in Slater determinant, harmonic oscillator bases, and we show how to
  remove spurious center of mass components exactly in order to derive
  translationally-invariant overlap integrals. We study in particular
  the Borromean \nuc{6}{He} nucleus using realistic chiral nuclear
  interactions, and we demonstrate that the observed clusterization in
  this system is a Pauli focusing effect. The inclusion of three-body
  forces has a small effect on this structure. In addition, we discuss
  the issue of absolute normalization for spectroscopic factors, which
  we show is larger than one. As part of this study we also perform
  extrapolations of ground-state observables and investigate the
  dependence of these results on the resolution scale of the
  interaction.
\end{abstract}


\maketitle

\section{Introduction} 
%
The structure of light nuclei is a very rich subject. A particularly
interesting phenomenon is the importance of clusterization. Cluster
structures appear frequently around reaction thresholds, and are
manifested, e.g., in large cluster form factors~\cite{austern1970}. It
is often modeled assuming that intrinsic cluster degrees of freedom are
frozen, thus reducing the full many-body problem to an effective
few-body one. However, the appearance of clusterization from a
microscopic perspective remains to be
elucidated~\cite{Ikeda:1968io,wildermuth1977,Okoiowicz:2012eu}.
It is not clear how strong, short-range correlations, induced by realistic
nuclear forces, propagate to longer-range cluster structures. In
addition, given the fermionic nature of nucleons we can be sure that
antisymmetrization at the many-body level will always play an important
role.

For light nuclei we have seen major progress in the development of
\emph{ab initio}
approaches~\cite{2013PrPNP..69..131B,Pieper:489644,Hagen:2010dd}. The
state-of-the-art methods are based on controlled approximations and the
underlying computational schemes account for successive many-body
corrections in a systematic way~\cite{2013PrPNP..68..158L}. Recently,
the application of chiral effective field theory
(EFT)~\cite{Ordonez:1996em,Bedaque:2002gm,Machleidt:2011gh,Epelbaum:2009hy}
and renormalization-group techniques~\cite{Bogner:2010ih} has resulted
in a systematic approach to the nuclear interaction.

Even more recently, \emph{ab initio} approaches began to bridge the gap from
nuclear structure to
reactions~\cite{Nollett:2007fb, Quaglioni:2008bb, 2010PhRvC..82a4310J,
  Forssen:2013eh}. 
Direct reactions, such as stripping and pickup of a single nucleon,
constitute a current frontier for these methods. In contrast, there
exists a rather standard approximation to treat such reactions within
phenomenological models~\cite{2003ARNPS..53..219H}, that uses
spectroscopic factors as input
parameters~\cite{Macfarlane:1960er,Bang:1985vi,austern1970}. The
spectroscopic factor corresponds to the integrated norm of the cluster
form factor. From a microscopic perspective it is a purely theoretical
construct that is defined from wave function overlaps. It is expected
that a correct treatment of translational invariance will be important
for this quantity. The particular case of three-body channels in
\nuc{6}{He} was studied by Timofeyuk~\cite{Timofeyuk:2001fd} who found a
significant increase of the normalization when using a
translation-invariant shell model. This observation was verified by
Brida~\cite{Brida:2010jp} using a microscopic Monte Carlo approach with
schematic interactions.

In this Paper we derive algebraic expressions for calculating
translationally invariant cluster form factors from no-core shell model
(NCSM) wave functions. We restrict ourselves to three-body
$\mathrm{core}+N+N$ channels, and will apply our formalism to study
$\nuc{4}{He}+n+n$ cluster structures in the Borromean system
\nuc{6}{He}~\cite{Zhukov:1993gf}. Despite its short beta-decay lifetime,
a series of precision measurements on \nuc{6}{He} ground state
properties have recently been performed. Its binding energy was measured
using the TITAN Penning trap mass spectrometer~\cite{Brodeur:2011kn},
and its charge radius was determined from laser
spectroscopy~\cite{Wang:2004jk,Mueller:2007ja}. In addition, there are
several theoretical studies of \nuc{6}{He} in the literature. From inert
cluster models~\cite{Zhukov:1993gf,Papadimitriou:2011ik} and microscopic
methods~\cite{2004NuPhA.738..357N, Baye:1994fv, Korennov:2004bq,
  Brida:2010jp} with phenomenological or semi-realistic interactions, to
\emph{ab initio} approaches~\cite{Caurier:2006p197, Bacca:2012ku,
  2008NCimR..31..709P, 2013PhRvC..88c4320Q} using high-precision nuclear
interactions.

The structure of the paper is the following. In Sec~\ref{sec:theory} we
give a brief introduction to the NCSM and we present the derivation of
algebraic expressions for $\mathrm{core}+N+N$ channel form factors.
In Sec.~\ref{sec:results} we present our results for ground-state
properties of \nuc{6}{He} using chiral interactions. Then we turn to the
overlap of \nuc{6}{He} with $\nuc{4}{He} + n + n$. We plot the
correlation density and decompose the cluster form factor into different
components of a hyperspherical harmonics expansion.
A discussion of our results is presented in Sec.~\ref{sec:discussion}
where we also give an outlook.
\section{Theoretical formalism%
\label{sec:theory}}
%
In the NCSM we consider a system of $A$ point-like non-relativistic
nucleons. The many-body basis is constructed from Slater determinants
(SD) of harmonic oscillator (HO) single-particle states. A basis
truncation is introduced by including all HO configurations up to a
certain energy cutoff (defined by the parameter \nm). This particular
choice of basis truncation guarantees translational invariance as all
eigenstates will factorize into a product of a state depending on
intrinsic coordinates and a state depending only on the center-of-mass
(CM) coordinate.  Eigensolutions with spurious CM excitations can then
be shifted up in the spectrum by adding a Lawson projection
term~\cite{1974PhLB...53..313G} to the Hamiltonian.
The NCSM Hamiltonian contains realistic two- and three-body nuclear
interactions, and the resolution scale of the Hamiltonian matrix is
usually lowered with similarity transformations. See, e.g.,
Ref~\cite{2013PrPNP..69..131B} for a more detailed description of the
NCSM method.

Since we want to compute a translationally invariant cluster form
factor, but still work with wave functions expressed in the NCSM basis
with single-particle coordinates, we need to take special care to remove
spurious CM components. A framework for performing this was introduced
in~\cite{Navratil:2004en} and is here generalized to the case of
three-body $\mathrm{core}+N+N$ channels.
\subsection{Coordinates and three-body cluster wave function}
We define the following set of Jacobi coordinates for an $A$-nucleon
system, adopting the notation of~\cite{Navratil:2004en}, where $\vec{\xi}_0$
is the $A$-body CM coordinate and $\boldsymbol{\xi}$
is the set of
normalised Jacobi coordinates for the $A-2$ particles in the core. The
relative coordinates for the clusters are defined as
\begin{subequations}
  \label{eq:coordinates}
  \begin{align}
    \vec{\eta}=&
    \sqrt{\frac{2(A-2)}{A}}\Bigg[\frac{1}{A-2} \sum_{i=1}^{A-2} \vec{r}_{i}
   -\frac{1}{2}(\vec{r}_{A-1}+\vec{r}_{A})\Bigg],\\
    \vec{\nu}=&\sqrt{ \frac{1}{2} } [ \vec{r}_{A-1} -\vec{r}_A ], 
  \end{align}
\end{subequations} 
which correspond to the normalised ``T-coordinate'' system of
$\mathrm{core}+N+N$.
%
Using this set of Jacobi coordinates we define cluster-separated
$A$-nucleon wave functions, that will be equivalent to a basis set in
the continuous variables $(\eta,\nu)$. Each basis function corresponds
to a set of frozen relative distances, which is reflected by two Dirac
$\delta$-functions
\begin{widetext}
\begin{align}
 \begin{split}
  &
  \braket{\boldsymbol{\xi}\eta'\vec{\eta}\nu'\vec{\nu}\boldsymbol{\sigma}
    \boldsymbol{\tau}}
  {\Phi^{AJMTM_T}_{{\alpha}{M}_{\alpha}};\delta_{\eta}\delta_{\nu} } 
    =\sum (l_{\eta}m_{\eta}l_{\nu}m_{\nu}|LM_L)(I_2M_2I_3M_3|I_{23}M_{23}) \\
&\times (I_1M_1I_{23}M_{23}|SM_S)  (LM_LSM_S|JM)
    (T_2M_{T_2}T_3M_{T_3}|T_{23}M_{T_{23}})(T_1M_{T_1}T_{23}M_{T_{23}}|TM_T) \\
   &\times  \frac{\delta(\eta-\eta')}{\eta\eta'}\frac{\delta(\nu-\nu')}{\nu\nu'} Y_{l_{\eta}m_{\eta}}(\hat{\eta}) Y_{l_{\nu}m_{\nu}}(\hat{\nu}) \braket{\sigma_{A-1}\tau_{A-1}}{ I_2
        M_2T_2M_{T_2}} \\
&\times
 \braket{\boldsymbol{\xi},\sigma_1\ldots\sigma_{A-2},\tau_1\ldots\tau_{A-2}}{(A-2)\alpha_1 I_1M_1T_1M_{T_1}} ,
 \end{split}
  \label{eq:3wf-dirac}
\end{align}
\end{widetext} 
where $\boldsymbol{\sigma}=\sigma_1, \ldots, \sigma_A$ and
$\boldsymbol{\tau}=\tau_1, \ldots, \tau_A$ are the spin and isospin
coordinates of the $A$ nucleons, and ${\alpha} \equiv \{\alpha_1 I_1T_1,
I_2T_2,I_3T_3; LS\}$ denotes the three-body channel in LS coupling with
the corresponding projection quantum numbers ${M}_{\alpha}$. The core,
with $A-2$ nucleons, has total angular momentum $I_1$ and isospin $T_1$,
while $\alpha_1$ correspond to additional quantum numbers needed to
characterize the eigenstate. Particles 2 and 3 are single nucleons so
$I_2=I_3=T_2=T_3=1/2$. The $A$-body system has a total angular momentum
$J$ and a total isospin $T$ with projections $M$ and $M_T$,
respectively.
%
\subsection{Three-body channel form factor}
The $\mathrm{core}+N+N$ three-body channel form factor for an $A$-body
state $\lambda J T$, with $\lambda$ denoting additional quantum numbers
necessary to characterize the state, can be defined as the overlap
integral
\begin{align}
\label{eq:overlap}
\begin{split}
&u^{A\lambda JT}_\alpha(\eta,\nu)
=\braket{A\lambda JT}{\mathcal{A}_{A-2,1,1} \Phi^{AJT}_\alpha;\delta_{\eta}\delta_{\nu} } \\
&\; =\sum_{n_{\eta},n_{\nu}}\sqrt{\frac{A!}{(A-2)!}}R_{n_{\eta}l_{\eta}}(\eta)R_{n_{\nu}l_{\nu}}(\nu)\\
&\; \times\braket{A\lambda JT}{\Phi^{AJT}_\alpha;n_{\eta}l_{\eta},n_{\nu}l_{\nu}},
\end{split}
\end{align}
where $\mathcal{A}_{A-2,1,1}$ is a cluster antisymmetrizer. It permutes
particles between the clusters, and gives a simple combinatorial factor
when acting on the fully antisymmetrized bra state. In this expression
we have expanded the Dirac $\delta$-functions in terms of radial HO
functions, $R_{nl}$, defined with the HO length parameter $b=\sqrt{\hbar
  / m\Omega}$, with $m$ the nucleon mass and $\Omega$ the HO
frequency. The new basis functions
$\braket{\boldsymbol{\xi}\vec{\eta}\vec{\nu} \boldsymbol{\sigma}
  \boldsymbol{\tau}}
{\Phi^{AJMTM_T}_{{\alpha}{M}_{\alpha}};n_{\eta}l_{\eta},n_{\nu}l_{\nu}}$
are identical to Eq.~\eqref{eq:3wf-dirac}, but with the delta functions
replaced by HO functions
$R_{n_{\eta}l_{\eta}}(\eta)R_{n_{\nu}l_{\nu}}(\nu)$.
The spectroscopic factor is the norm of the overlap integral. 

The relationship between an $A$-nucleon wave function expressed in the
SD basis, with $0S$ CM motion as guaranteed by the Lawson projection,
and the corresponding state in Jacobi coordinates is
\begin{align}
\begin{split}
&\braket{\vec{r}_1...\vec{r}_A \boldsymbol{\sigma} \boldsymbol{\tau}}{ A\lambda
JMTM_T}_\mathrm{SD}\\
&=\braket{\boldsymbol{\xi} \vec{\eta} \vec{\nu} \boldsymbol{\sigma} \boldsymbol{\tau}}{A\lambda
JMTM_T}\psi_{000}(\vec{\xi}_0).
\end{split}
\end{align}
By applying this relationship to the composite and cluster states it is
possible to relate the overlap in Jacobi coordinates with an overlap
expressed in an SD basis
\begin{equation}
\label{eq:jacsd}
\begin{split}
& \braket{A\lambda
   JT}{\mathcal{A}_{A-2,1,1}\Phi^{AJT}_\alpha;n_{\eta}l_{\eta},n_{\nu}l_{\nu}}
\\ 
&\; = \frac{ {}_\mathrm{SD} \braket{A\lambda JT}{\mathcal{A}_{A-2,1,1}
  \Phi^{AJT}_\alpha;n_{\eta}l_{\eta},n_{\nu}l_{\nu}}{}_\mathrm{SD}}
{\braket{n_{\eta}l_{\eta}00l_{\eta}}{00n_{\eta}l_{\eta}l_{\eta}}_{\frac{2}{A-2}}},
\end{split}
\end{equation}  
where the denominator is the general HO bracket~\cite{Trlifaj:1972jv}
that results from a Talmi-Moshinksky transformation.  A second
transformation takes us from $\eta$ and $\nu$ to single-particle
coordinates (subscripts $a$ and $b$) for the two nucleons outside the
core. Finally, recoupling spins and integrating over the intrinsic
coordinates we arrive at an expression for the form factor expressed in
terms of double-reduced matrix elements between SD eigenstates
\begin{widetext}
\begin{align}
\begin{split}
\label{eq:sd-overlap}
&u^{A\lambda JT}_\alpha(\eta,\nu) 
=\sum_{\substack{n_{\eta}l_{\eta}\\ n_{\nu}l_{\nu}}\ldots}\frac{R_{n_{\eta}l_{\eta}}(\eta)R_{n_{\nu}l_{\nu}}(\nu)}{\braket{n_{\eta}l_{\eta}00l_{\eta}}{00n_{\eta}l_{\eta}l_{\eta}}_{\frac{2}{A-2}}}
(-1)^{3I_1+I_{23}+J_{ab}-T_{23}-S+L}
\\
&\times  \braket{n_al_an_bl_bL}{n_{\eta}l_{\eta}n_{\nu}l_{\nu}L}_{1} \frac{\hat{L}\hat{S}\hat{J}_{ab}^2\hat{j}_a\hat{j}_b}{\hat{J}\hat{T}} \sexj{L}{I_{23}}{J_{ab}}{I_1}{J}{S}\nioj{l_a}{l_b}{L}{I_3}{I_2}{I_{23}}{j_a}{j_b}{J_{ab}}\\
&\times {}_{\mathrm{SD}}\rrmatrixel{A\lambda JT}{\left[a_{n_al_aj_at_a}^\dagger
    a_{n_bl_bj_bt_b}^\dagger\right]^{J_{ab}T_{ab}}}{(A-2)\alpha_1 I_1T_1}{}_{\mathrm{SD}},
\end{split}
\end{align}
\end{widetext}
where $J_{ab}$ ($T_{ab}$) is the coupled total spin (isospin) of the two nucleons.
The $a^\dagger a^\dagger$ matrix elements are calculable using a special
version of our transition density code~\cite{Navratil:2004en}.
%
\section{Results%
\label{sec:results}}
%
We will start this section with a presentation of our results for
\nuc{6}{He} ground-state observables. The second part will then be
devoted to the extraction of three-body channel form factors and the
corresponding spectroscopic factors. Our calculations are performed in
the NCSM for model spaces up to $\nm=16$, corresponding to a basis
dimension of $3.6 \cdot 10^8$. Unless otherwise stated, we employ the
Idaho chiral \nn\ interaction at N${}^3$LO with a 500 MeV regularization
cutoff~\cite{Machleidt:2011gh}. The interaction is evolved in the
two-body free space using the SRG flow equation~\cite{Bogner:2010ih} in
order to compute a phase-shift equivalent, effective two-body
interaction. We note that the truncation of the evolution at two-body
level will impose a violation of formal unitarity for the transformation
in the many-body space. A specific aim of this study is therefore to
investigate the dependence of our results on the SRG flow parameter
\lsrg. We will use a physically motivated range of resolution scales,
corresponding to $\lsrg = 1.8-2.2 \mathrm{~fm}^{-1}$.

\begin{figure}[thb]
\centering
\includegraphics[width=1\columnwidth]{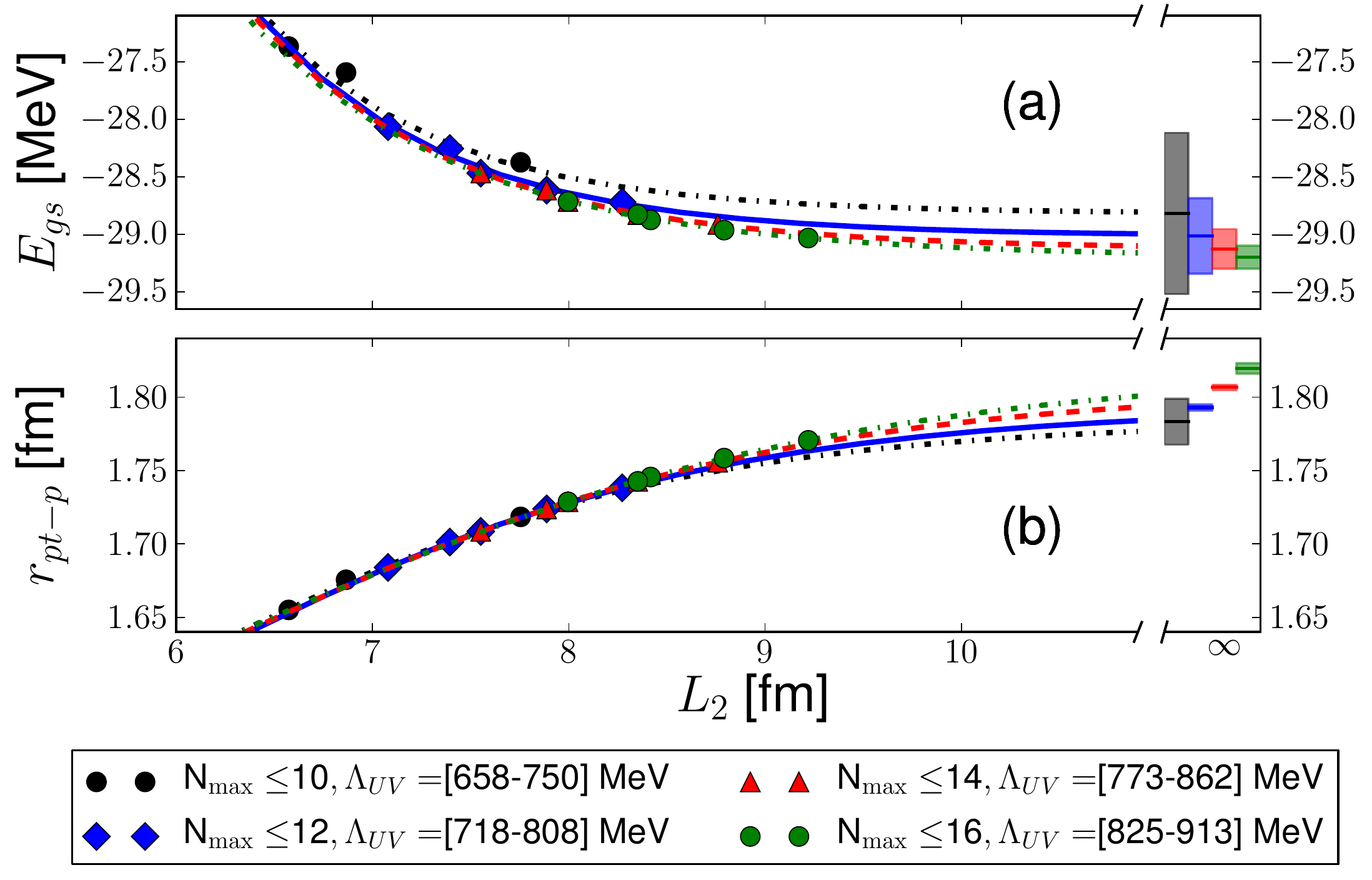}
\caption{\co\ Extrapolation of \nuc{6}{He} binding energy (a) and
  point-proton radius (b) as a function of the IR cutoff
  parameter. Results are obtained with the NCSM using an SRG-evolved
  chiral two-body interaction ($\lsrg = 2.0$~fm${}^{-1}$). See text for
  details on the extrapolation procedure.%
\label{fig:6he-energiesL}}
\end{figure}

For any choice of realistic interaction, the ground-state energy of a
many-body system calculated in a truncated space shows a dependence on
the basis parameters \nm\ and \ho. By construction, our results should
be independent of \ho\ in the limit of infinite model space. We will now
discuss extrapolations of our \nuc{6}{He} finite-space results.
Our oscillator basis truncation can be translated into corresponding
infrared (IR) and ultraviolet (UV)
cutoffs~\cite{Coon:2012dl,Furnstahl:2012eu,More:2013kf}. Following
Refs.~\cite{Furnstahl:2012eu,More:2013kf}, we define the UV momentum
cutoff $\Lambda_\mathrm{UV} = \sqrt{2(N+3/2)}\hbar/b$, where $N$ is the
truncation in the single-particle basis ($N=\nm+1$ for $p$-shell
nuclei). For the IR parameter we use $\Lambda_\mathrm{IR} = 1/L$, with
$L = L_2 \equiv \sqrt{2(N+3/2+2)} b$ as suggested in
Ref.~\cite{More:2013kf}.
Working in very large model spaces, we are able to capture the UV
physics of the softened interaction. As a consequence, the IR correction
will be the most important one and we use the IR dependence of the
energy that was derived in Ref.~\cite{Furnstahl:2012eu}
\begin{equation}
  E_L = E_\infty + A e^{-2 k_\infty L},
\end{equation}
where $k_\infty$ should be related to the binding momentum, but in
practice will be used as a free fit parameter together with $A$ and the
desired $E_\infty$. In addition, we use the suggested IR correction
formula for the point-proton radius~\cite{Furnstahl:2012eu}
\begin{equation}
\langle r^2 \rangle_L \approx \langle r^2 \rangle_\infty \left[ 1 - (c_0
  \beta^3 + c_1 \beta ) e^{-\beta} \right],
\end{equation}
where $\beta \equiv 2 k_\infty L$. We use $c_0$, $c_1$, and $\langle r^2
\rangle_\infty$ as fit parameters, but keep $k_\infty$ fixed from the
energy fit. Our calculated data for these two ground-state observables
are presented in Fig.~\ref{fig:6he-energiesL} together with the
extrapolation curves. 

In practice, we want to test the performance of the extrapolation
procedure as a function of the model space truncation. We start by
imposing a rather small $\nm$ truncation and collect the results
computed at the largest values of $\Lambda_\mathrm{UV}$ into one data
set of five points that is used for the curve fit. The error bar
reflects the variance from the least-squares fit. It does not include an
estimate of the systematic error from the extrapolation. This procedure is
then repeated for increasing $\nm$, i.e.\ including more data, until we
finally use the unrestricted data set. The evolution of extrapolated
result with error bars from $\nm \le$10 to 16 is shown in
Fig.~\ref{fig:6he-energiesL}. While we find a consistent set of results
for the energies, we note that there is a trend of increasing
point-proton radius that calls for further investigation.
Final results, obtained with the unrestricted data set $\nm \le 16$,
are presented in Table~\ref{tab:6he-energy} for three different SRG
parameters. The fact that the unitarity of the transformation is only
approximate leads to binding and separation energy variations of a few
hundred keV, but is hardly noticable within error bars of the
extrapolated radius.
\begin{table*}
  \caption{\label{tab:6he-energy} Extrapolated results for the \nuc{6}{He}
    binding energy, two-neutron separation energy, and point-proton
    radius. Results are obtained with the NCSM using an SRG-evolved chiral
    two-body interaction with three different SRG flow parameters. The
    last column shows theoretical results from Bacca et
    al.~\cite{Bacca:2012ku} using the hyperspherical-harmonics approach
    with the same chiral interaction, but employing the
    $V_{\mathrm{low}-k}$ renormalization technique.}
\begin{ruledtabular}
\begin{tabular}{c d  d d d d}
& \multicolumn{1}{c}{Exp.~\cite{Brodeur:2011kn}} 
& \multicolumn{3}{c}{This work}
&\multicolumn{1}{c}{Bacca \emph{et al.}~\cite{Bacca:2012ku},\cite{Bacca:2009jf}}  \\
& & \multicolumn{1}{c}{$\lsrg = 1.8$} 
& \multicolumn{1}{c}{$\lsrg = 2.0$}
& \multicolumn{1}{c}{$\lsrg = 2.2$} 
&\multicolumn{1}{c}{$V_{\mathrm{low}-k}$ ($\Lambda=2.0$~fm${}^{-1}$})\\
\hline
$E_\mathrm{gs}$ [MeV] & 29.269 & 29.67(3) & 29.20(11) & 28.61(22)&  29.47(3)\\
$S_{2n}$ [MeV] & 0.975 &1.22(2) & 0.95(10) &0.68(22)  &0.82(4) \\
$r_\mathrm{pt-p}$ [fm] & 1.938(23) &1.820(4)& 1.820(4)& 1.815(8) &1.804(9)\\
\end{tabular} 
\end{ruledtabular}
\end{table*}

\begin{figure}[thb]
\centering
\includegraphics[width=\columnwidth]{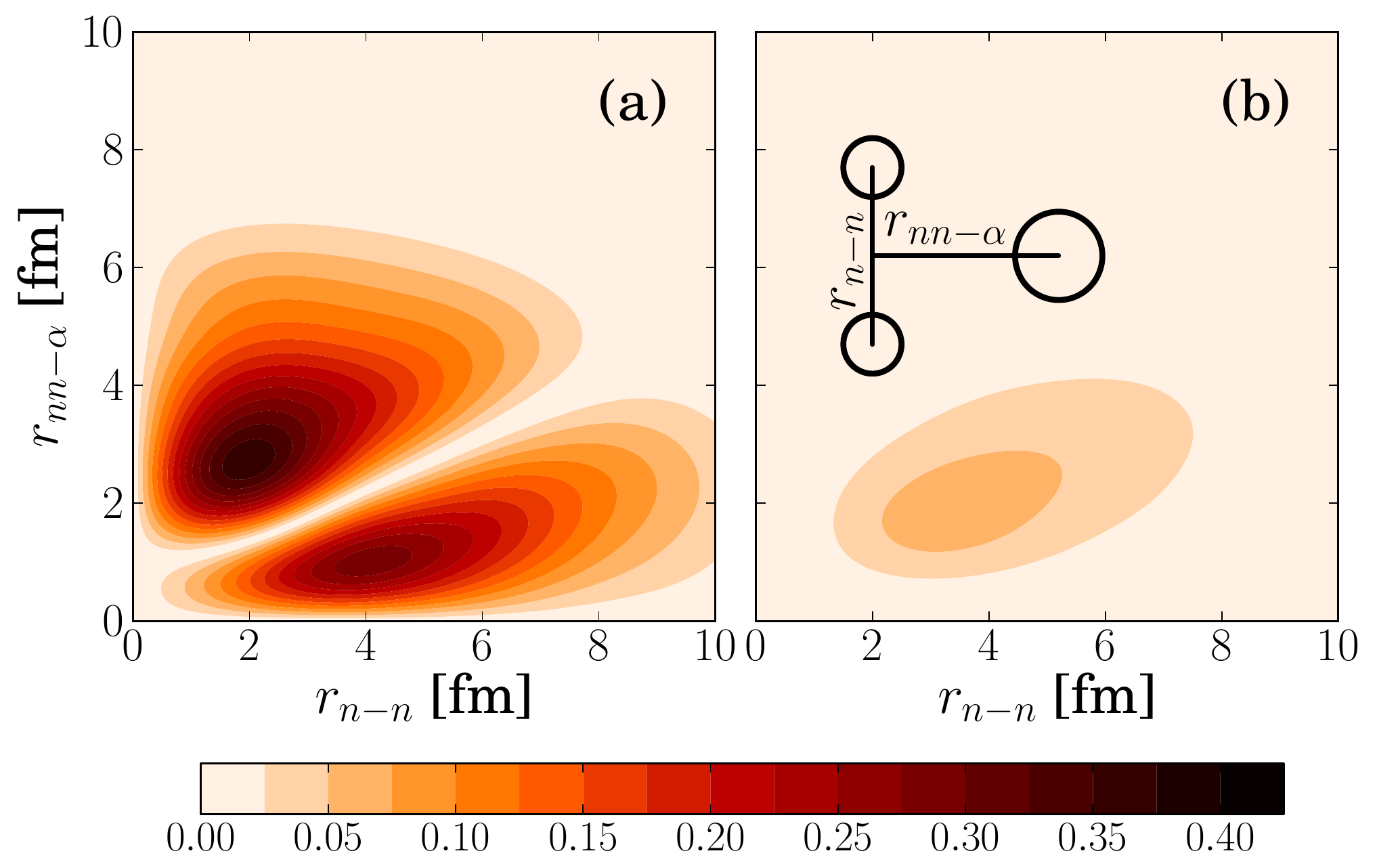}
\caption{\co\ Contour plots of the translationally-invariant  three-body channel form factor
  $\left\langle \nuc{6}{He}(0^+) | \nuc{4}{He}(0^+) + n + n
  \right\rangle$ calculated from NCSM wave functions. The two allowed
  channels for this overlap, $L=S=0$ and $L=S=1$, are shown in panels
  (a) and (b), respectively.%
\label{fig:6he-overlap}}
\end{figure}

Next we turn to the computation of three-body channel form factors from
our microscopic wave functions. We employ Eq.~\eqref{eq:sd-overlap} with
NCSM wave functions up to $\nm=14$. In Fig.~\ref{fig:6he-overlap} we
show the main ($L=S=0$) component of the $\left\langle \nuc{6}{He}(0^+)
  | \nuc{4}{He}(0^+) + n + n \right\rangle$ overlap, and the much
smaller $L=S=1$ component. Our microscopic calculation provides a
beautiful confirmation of the two-peak structure of this form factor, as
reported in earlier phenomenological cluster model
studies~\cite{1986NuPhA.453..365K, Zhukov:1993gf, Papadimitriou:2011ik},
and within a microscopic model with schematic
interactions~\cite{Brida:2010jp}. It is clear that the small $L=S=1$
component does not show any signs of a similar structure.

Now we are uniquely positioned to analyse this form factor behavior and
to understand the origin of the observed clustering. To begin with, we
note that the so called di-neutron configuration has the largest peak
probability. This is expected since it contributes to a shift of the
position of the charged core with respect to the total center-of-mass,
and therefore to an increased charge radius, which is consistent with
experimental findings. Note, however that the average distance between
the two neutrons is not very small ($\sim 2$~fm), and it can be expected
that the influence on the charge radius will diminish when adding
additional neutrons. Indeed, it has been shown experimentally that the
charge radius increases for \nuc{6}{He}, but decreases again in
\nuc{8}{He}~\cite{Brodeur:2011kn}. Gamow shell model
calculations~\cite{Papadimitriou:2011ik}, that incorporate continuum
structures explicitly, confirm that the amplitude of the di-neutron
configuration is reduced when going from \nuc{6}{He} to \nuc{8}{He}.

\begin{figure}[thb]
\includegraphics[width=0.9\columnwidth]{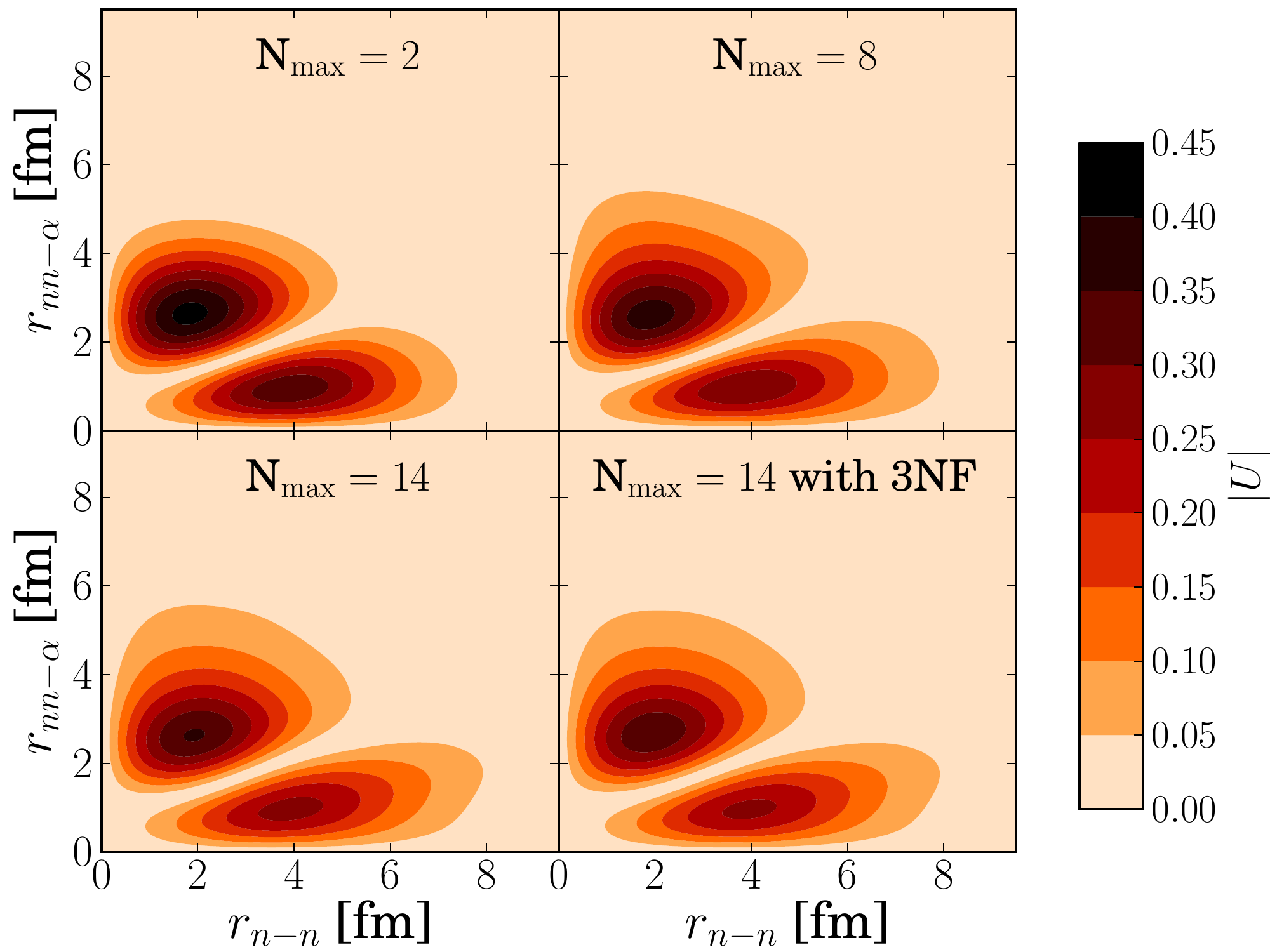}
\caption{\co\ 
Model-space dependence of the three-body channel form
factor $\left\langle \nuc{6}{He}(0^+) | \nuc{4}{He}(0^+) + n + n
  \right\rangle$. The main $L=S=0$ channels is shown for a sequence of
  calculations with increasing \nm\ performed with $\lsrg=2.0$ fm${}^{-1}$ and
  $\ho=16$ MeV. The lower right panel shows the same contour plot
  using an SRG-evolved chiral $\nn+\nnn$ interaction~\cite{roth:private}.%
\label{fig:6he-pauli}}
\end{figure}

We claim that the origin for the observed cluster structure in
\nuc{6}{He} is the Pauli principle. To substantiate this statement we
show in Fig.~\ref{fig:6he-pauli} a sequence of contour plots obtained at
$\nm = 2, 8, 14$. The same structure is clearly seen in all three
panels, although the former ones represent calculations that are far
from converged, i.e.\ the model space is much too small to accomodate
the correlations induced by the interaction. However, what is present
already at the smallest model spaces is the correct antisymmetrization.
The importance of a proper treatment of antisymmetrization was stressed
in previous microscopic studies, see
e.g.~Ref.~\cite{Brida:2010jp}. Obviously, this feature remains a very
weak point in models with inert clusters.
The wave function obtained with a three-body Hamiltonian (SRG-evolved
chiral $\nn+\nnn$~\cite{roth:private}) exhibits the same clusterization
structure, as can be observed in the lower right panel of
Fig.~\ref{fig:6he-pauli}.

In order to analyse the cluster form factors further we have performed a
projection on hyperspherical harmonics (HH) basis functions. The
hypercoordinates $(\rho, \theta, \hat{\eta}, \hat{\nu})$ are related to
the Jacobi coordinates $(\vec{\eta}, \vec{\nu})$ via
$\eta=\rho\cos\theta$, 
$\nu = \rho\sin\theta$.
The cluster form factor from Eq.~\eqref{eq:sd-overlap}, expressed in
$(\rho,\theta)$ coordinates, can be projected onto the HH basis
\begin{align}
\begin{split}
u^{A\lambda JT}_\alpha(\theta,\rho)
&=\frac{1}{\rho^{5/2}} \sum_{K,l_{\eta},l_{\nu}} 
\chi^{A\lambda JT}_{\alpha, K l_{\eta} l_{\nu}} (\rho)
\psi_K^{l_{\eta},l_{\nu}}(\theta), \mathrm{~where} \\
\chi^{A\lambda JT}_{\alpha, K l_{\eta} l_{\nu}} (\rho) 
&= \rho^{5/2} \int_0^{\frac{\pi}{2}}\mathrm{d}\theta'\sin^2\theta'\cos^2\theta'
\psi_K^{l_{\eta},l_{\nu}*}(\theta')\\
\times\sum_{n_{\eta},n_{\nu}}
&C^{A\lambda JT}_{\alpha, n_{\eta}l_{\eta},n_{\nu}l_{\nu}}R_{n_{\eta}l_{\eta}}(\theta',\rho)R_{n_{\nu}l_{\nu}}(\theta',\rho),
\label{eq:hhproj}
\end{split}
\end{align}
where $\psi_K^{l_{\eta},l_{\nu}}(\theta)$ is the
hyperangular basis function~\cite{FabredelaRipelle1983281}, and the last row 
is a compact formulation of the RHS of Eq.~\eqref{eq:sd-overlap}.  

\begin{figure}[thb]
\centering
\includegraphics[width=\columnwidth]{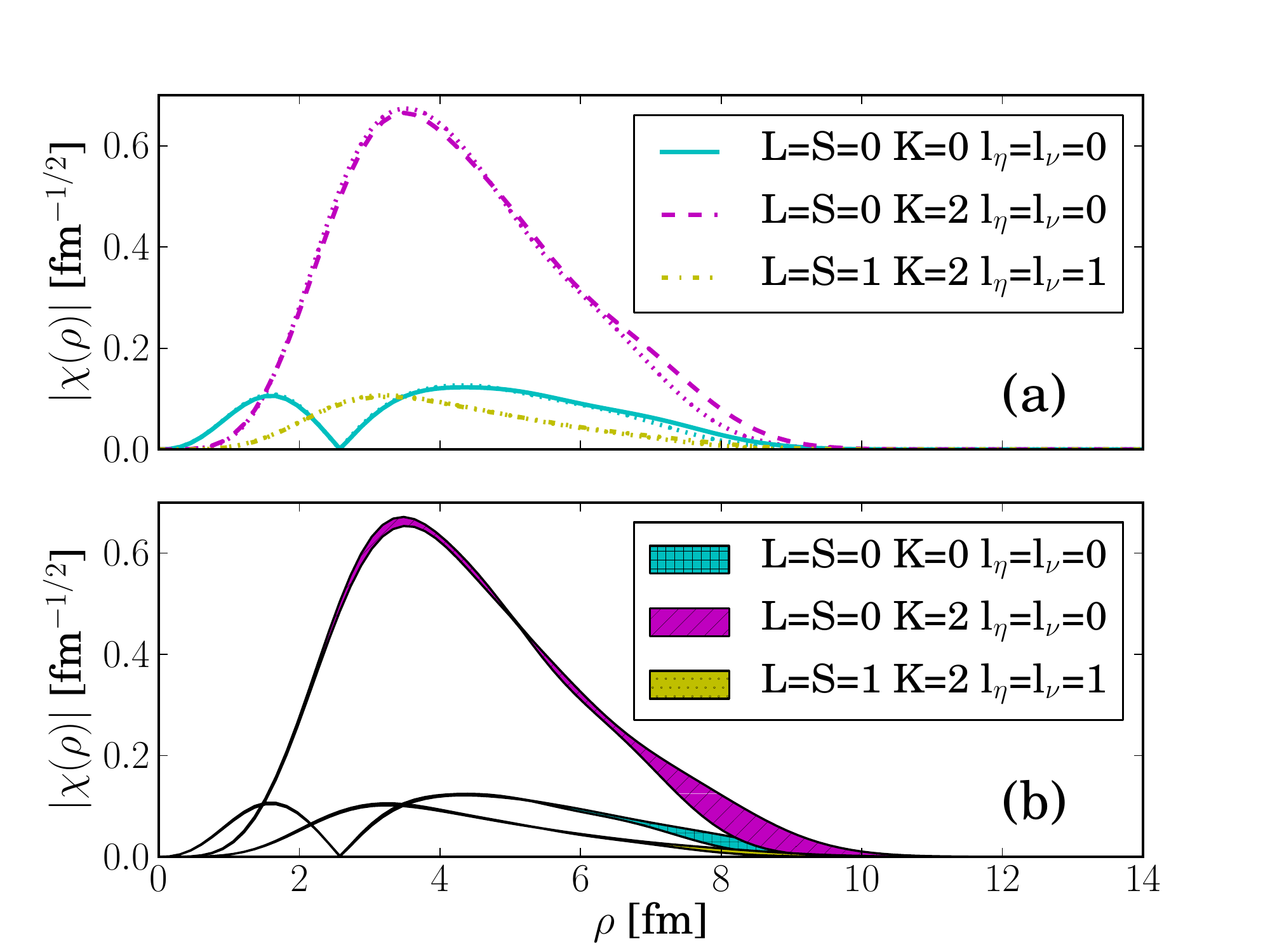}
\caption{\co\ Hyperradial functions obtained from NCSM three-body
  channel form factors calculated with an SRG-evolved chiral two-body
  interaction ($\lsrg=2.0$ fm${}^{-1}$). Panel a: N$_{\mathrm{max}}$
  dependence for a fixed frequency, $\hbar\Omega=20 $ MeV. Thick lines
  correspond to $\nm=14$ results, while dotted ones
  are N$_{\mathrm{max}}$=12. Panel b: Shaded bands correspond to a
  fixed model space ($\nm=14$) and a range of HO frequencies, $\hbar\Omega=[16,22]$. %
  \label{fig:6he-hh}}
\end{figure}

In Fig.~\ref{fig:6he-hh} we show the hyperradial functions for the three
most important terms of this basis expansion. We focus in particular on
the sensitivities to variations in the model space (panel a) and HO
frequency (panel b). The interior part of the overlap is very well
converged. However, we can observe that the expected exponential
tail~\cite{Zhukov:1993gf} is not reproduced. Increasing \nm\ we find
that the tail builds up slowly.  A similar behavior is found when
varying the HO frequency, as can be seen from the hatched bands in
Fig.~\ref{fig:6he-hh}b. Small frequencies correspond to large oscillator
lengths, and therefore reproduce longer tails. However, a proper
treatment of the continuum and long-range asymptotics is needed to
describe this region more accurately. Such work is ongoing in the
framework of the NCSM/RGM~\cite{2013PhRvC..88c4320Q}.
Note that the total norms of the hyperradial functions are determined
mainly by the amplitude in the internal region, where the dependence on
\nm\ and \ho\ is small. The relative weights of the five main
components, as well as the total sum (i.e., the spectroscopic factor),
are presented in Table~\ref{tab:6he-hh2}. The calculations are performed
in the $\nm=14$ model space for various SRG parameters and HO
frequencies. We find that the sensitivity to variations in HO frequency
and \nm\ is $\lesssim 1$~\%.
We note that we have very small variations around a total spectroscopic
factor of $\sim 1.3$. The wave function obtained using a chiral
$\nn+\nnn$ Hamiltonian~\cite{roth:private} displays an HH expansion that
is qualitatively very similar.
We have also computed the overlap with an excited core, $\left\langle
  \nuc{6}{He}(0^+) | \nuc{4}{He}(2^+) + n + n \right\rangle$. This
channel corresponds to the first excited $2^+$ state in the NCSM,
situated just below the $2n+2p$ threshold but with slow energy
convergence. Using $\lsrg = 2.0$~[fm]$^{-1}$, we find the total
spectroscopic factor 0.30(4).
\begin{table*}
\caption{\label{tab:6he-hh2} Relative weights (in \%) of the HH
  expansion terms for the
  three-body channel form factor $\left\langle \nuc{6}{He}(0^+) |
  \nuc{4}{He}(0^+) + n + n \right\rangle$ calculated from NCSM wave
  functions. The last row shows the total spectroscopic
  factor.} 
\begin{ruledtabular}
\begin{tabular}{l c c| c c c c c c |c |c}
  \multicolumn{3}{c|}{Three-body channel} & \multicolumn{5}{c}{This work
    (\lsrg\ [fm${}^{-1}$],\ho\ [MeV])} & with 3NF & Ref.~\cite{Zhukov:1993gf} &
  Ref.~\cite{Brida:2010jp}  \\ 
  $K$ &$l_{\eta}=l_{\nu}$  &$L=S$ & $(1.8,20)$ & $(2.0,16)$ & $(2.0,20)$ &
  $(2.0,22)$ & $(2.2,20)$ &$(2.0,16)$& (cluster) &
  (microscopic) \\
  \hline
  0 & 0 & 0& 4.2 & 4.4 &4.3& 4.2 & 4.3&4.1 &4.2 & 4.0\\
  2 & 0 & 0&92.0& 91.7&91.9& 92.1& 92.0&91.3 & 82.1 & 79.9 \\
  2 & 1 & 1&  2.1& 2.1 & 2.2&2.2& 2.2& 3.0 & 11.2& 13.3\\
  6 & 2 & 0 & 1.1  & 1.2&1.1& 1.0&1.0& 1.0&1.7&1.9\\
  6 & 3 & 1 & 0.1  &0.1 &0.1& 0.1&0.1 & 0.1 & 0.8 &0.8\\
  \hline
  \multicolumn{3}{r|}{Spectroscopic factor:} & 1.3441&1.3263 &1.3340&
  1.3391 &1.3278& 1.3284&0.9851 &1.3957 \\ 
\end{tabular} 
\end{ruledtabular}
\end{table*}

\section{Conclusion and discussion%
\label{sec:discussion}}
In this paper we have derived expressions for translationally-invariant
$\mathrm{core}+N+N$ overlap integrals starting from microscopic wave
functions. We have used these overlap integrals to perform a microscopic
investigation of the clustering of \nuc{6}{He} into
$\nuc{4}{He}+n+n$. Large-scale \emph{ab initio} NCSM calculations were
performed with realistic nuclear interactions obtained from chiral
perturbation theory.
In addition, we used an SRG evolution to lower the resolution scale of
the many-body Hamiltonian. We generated a series of such interactions,
labeled by the SRG flow parameter, connected to each other by
(approximately) unitary transformations. All of them reproduce the same
two-nucleon, low-energy observables, but they have different
high-momentum properties. This implies a resolution-scale dependence in
the short-ranged part of the wave function. Therefore, we stress that
the overlap integrals (and their norms, the spectroscopic factors) are
not physical observables~\cite{2002PhLB..531..203F,Furnstahl:2010hg}. In
this work we have computed spectroscopic factors, as well as
ground-state observables, using our series of Hamiltonians connected by
SRG transformations.

In particular, we have studied the ground-state energy, two-neutron
separation energy, and point-proton radius of \nuc{6}{He}. We employed
published extrapolation schemes~\cite{Furnstahl:2012eu,More:2013kf} to
correct for the finite HO model space that were used in the
computations, and we found a very consistent set of results for the
energy while the extrapolation behavior of the radius calls for further
studies.

Concerning the cluster structure of the \nuc{6}{He} ground state we have
found that the total spectroscopic factor for $\left\langle
  \nuc{6}{He}(0^+) | \nuc{4}{He}(0^+) + n + n \right\rangle$ is
significantly larger than one. In contrast, phenomenological cluster
models assume that this quantity is normalized to unity (see e.g.\ the
results of Ref.~\cite{Zhukov:1993gf} in Table~\ref{tab:6he-hh2}). Our
result is consistent with the translation-invariant shell-model upper
limit of $25/16 \approx 1.56$ by
Timofeyuk~\cite{Timofeyuk:2001fd}. Furthermore, we have studied the \nm\
dependence of the overlap and could conclude that the clusterization is
clearly driven by the Pauli principle. By performing an HH expansion of
our three-body channel form factor, we found a strong dominance of the
$K=2$, $L=S=0$ channel, which drives a two-peak structure in the
$T$-system of Jacobi coordinates. Furthermore, we can note that
phenomenological approaches predict a significantly enhanced $L=S=1$
channel as compared to our NCSM results with realistic, chiral
interactions. We also find a non-negligible overlap with an excited
core $\nuc{4}{He}(2^+)$.

This work clears the path for further investigations of three-body
clustering in light nuclei. Through the microscopically extracted
overlap integrals we have a natural interface with reaction calculations
that build on cluster degrees of freedom. In addition, we can combine
this work with the ongoing development of the
NCSMC~\cite{2013PhRvC..88c4320Q}. The latter aims to couple NCSM $A$-body
eigenstates with \emph{ab intio} cluster wave functions. The formalism
presented in this paper will allow a detailed investigation of the
resulting cluster structures.
\begin{acknowledgments}
  The research leading to these results has received funding from the
  European Research Council under the European Community's Seventh
  Framework Programme (FP7/2007-2013) / ERC grant agreement
  no.~240603. This work was also supported by the Swedish Research
  Council (dnr.~2007-4078). This research used computational resources
  at Chalmers Centre for Computational Science and Engineering (C3SE)
  provided by the Swedish National Infrastructure for Computing (SNIC). 
  We are much indebted to R. Roth for providing wave functions from the
  IT-NCSM and to Petr Navr\'atil, R. Wiringa, and N. Timofeyuk for insights and
  stimulating discussions on clustering in light nuclei.
\end{acknowledgments}

\bibliography{two-body-overlap,two-body-overlap_temp}
\bibliographystyle{apsrev}

\end{document}